# Spectroscopy of the Fractal Hofstadter Energy Spectrum

Kevin P. Nuckolls[1,2,a,*], Michael G. Scheer[2,*], Dillon Wong[1,2,*], Myungchul Oh[1,2,b,*], Ryan L. Lee[1,2], Jonah Herzog-Arbeitman[2], Kenji Watanabe[3], Takashi Taniguchi[4], Biao Lian[2], Ali Yazdani[1,2,‡]

[1] *Joseph Henry Laboratories, Princeton University, Princeton, NJ 08544, USA*

[2] *Department of Physics, Princeton University, Princeton, NJ 08544, USA*

[3] *Research Center for Functional Materials, National Institute for Materials Science, 1-1 Namiki, Tsukuba 305-0044, Japan*

[4] *International Center for Materials Nanoarchitectonics, National Institute for Materials Science, 1-1 Namiki, Tsukuba 305-0044, Japan*

[a] *Present Address: Department of Physics, Massachusetts Institute of Technology*

[b] *Present Address: Department of Semiconductor Engineering, POSTECH*

\* These authors contributed equally to this work.

‡ Corresponding author email: yazdani@princeton.edu

**Hofstadter's butterfly, the predicted energy spectrum for non-interacting electrons confined to a two-dimensional lattice in a magnetic field, is one of the most remarkable fractal structures in nature[1]. At rational ratios of magnetic flux quanta per lattice unit cell, this spectrum shows self-similar distributions of energy levels that reflect its recursive construction. For most materials, Hofstadter's butterfly is predicted under experimental conditions that are unachievable using laboratory-scale magnetic fields[1–3]. More recently, electrical transport studies have provided evidence for Hofstadter's butterfly in materials engineered to have artificially large lattice constants[4–6], such as those with moiré superlattices[7–10]. Yet to-date, direct spectroscopy of the fractal energy spectrum predicted by Hofstadter nearly 50 years ago has remained out of reach. Here we use high-resolution scanning tunneling microscopy / spectroscopy (STM / STS) to probe the flat electronic bands in twisted bilayer graphene near the predicted second magic angle[11,12], an ideal setting for spectroscopic studies of Hofstadter's spectrum. Our study shows the fractionalization of flat moiré bands into discrete Hofstadter subbands and discerns experimental signatures of self-similarity of this spectrum. Moreover, our measurements uncover a spectrum that evolves dynamically with electron density, displaying phenomena beyond that of Hofstadter's original model due to the combined**

**effects of strong correlations, Coulomb interactions, and the quantum degeneracy of electrons in twisted bilayer graphene.**

In 1976, Hofstadter showed that a periodic potential restructures the Landau levels of a two-dimensional electronic system in a magnetic field into a fractal energy spectrum, called "Hofstadter's butterfly." This spectrum resembles a butterfly in a plot against energy and magnetic field and shows self-similar distributions of energy bands when the number of magnetic flux quanta per unit cell of the periodic potential ("flux per unit cell"; $\Phi/\Phi_0$) is a rational number $p/q$ ($p$ and $q$ are coprime integers)[1]. The $B$-fields required to achieve the Hofstadter regime for an atomic potential are not feasible ($B > 1000$ T) because of the scale of physical constants ($h/e \approx 4100$ T nm$^2$); however, artificial superlattices, as first achieved in microstructured semiconductor heterostructures[4–6], provide realistic platforms for realizing Hofstadter's butterfly. More recently, moiré materials have emerged as high-quality Hofstadter systems[13], as observed in lattice-mismatched heterostructures of graphene and hexagonal boron nitride (hBN)[7–10,14–17]. With a moiré wavelength $\lambda_m$ nearly 100 times the interatomic spacing, Hofstadter's butterfly influences transport observables at tens of tesla. Unfortunately, spectroscopic experiments at these $B$-fields are still unachievable, prohibiting direct access to the fractal energy spectrum itself.

Here we use scanning tunneling spectroscopy (STS) to probe the fractal Hofstadter energy spectrum in a homobilayer moiré material, twisted bilayer graphene (TBG). The advantage of this material is its extremely long-wavelength ($\lambda_m \approx 25$ nm), highly homogeneous superlattice potential that admits clear signatures of Hofstadter's butterfly at $B$-fields as low as 1 T. Investigating the Hofstadter spectrum as a function of energy, magnetic field, and carrier density enables active control over the complex interplay of many electronic effects beyond Hofstadter's simplified model. This allows us to disentangle these effects using spectroscopic information that is inaccessible to transport[8–10,14] and thermodynamic[18,19] probes typically used to study Hofstadter systems.

**Flat Bands Near the Second Magic Angle**

TBG sits on an hBN/SiO$_2$/Si substrate, twisted near the predicted second magic angle ($\theta$ = 0.63°, 0.57° for Device A1, A2), as resolved in STM topographs (Fig. 1a; see SI for large-scale images). Consistent with previous TEM studies, topographic images show strong lattice relaxation effects at small twist angles that enlarge (minimize) the area of AB / BA-stacked (AA-stacked) regions[20,21]. At these angles, the Bistritzer-MacDonald model[11] predicts a dense series of flat electronic bands (Fig. 1b; also see SI), with a set of ten particularly narrow moiré bands (orange-colored) near charge neutrality.

Fig. 1c shows the tunneling differential conductance d$I$/d$V$($V_s$, $V_g$) obtained at the center of an AA site at zero $B$-field (Device A2; $\theta$ = 0.57°, $\varepsilon$ = 0.07%), where $V_s$ and $V_g$ are the TBG sample bias and back gate voltages, respectively. Nearly the entire accessible density range shows sharp peaks in the local density of states (LDOS) that pin to the Fermi level ($E_F$) over extended density ranges. Fig. 1d shows a zoomed-in plot of d$I$/d$V$($V_s$, $V_g$) near $V_g$ = 0 and Fig. 1e shows d$I$/d$V$($V_s$) at charge neutrality displaying two sharp LDOS peaks, each less than 20 meV in width. Each peak represents a flat electronic band, as predicted at this twist angle (Fig. 1b; see SI for details). These flat bands (Figs. 1d,e) are nearly as narrow in energy as those observed near the first magic angle[22]. However, at zero $B$-field, surprisingly no correlation-driven energy gaps[23–26] were observed at partial fillings of any of these flat band, even when probed at lower temperatures ($T$ = 200 mK).

The advantages of studying the Hofstadter spectrum in second magic-angle TBG are twofold. First, the moiré wavelength is large ($\lambda_m \approx$ 25 nm; Fig. 1a), enabling detailed measurements of the fractal spectrum at low $B$-fields without the interference of high-$B$-field Coulomb gap features that suppress low-energy tunneling spectroscopy[27–30]. Even smaller twist angles / large moiré periodicities might be desired; however, substantial energy overlap between moiré bands would complicate observations at such angles. In contrast, Fig. 1f shows the zero-bias conductance d$I$/d$V$($V_s$ = 0, $V_g$) from Fig. 1c. Particularly near the second magic angle, TBG shows many flat-band peaks that are relatively well-separated in density, as evidenced by deep suppressions of d$I$/d$V$($V_s$ = 0, $V_g$) near $v$ = 4$N$ for integer $N$. Therefore, TBG is a model system for studying Hofstadter's butterfly, striking a balance where nearly separated flat bands can cleanly split into Hofstadter subbands at low $B$-fields.

**Spectroscopy of Hofstadter Gaps**

To probe the Hofstadter spectrum of TBG, we performed STS measurements in a perpendicular $B$-field. Hofstadter's calculations[1] predict that at $\Phi/\Phi_0$ = 1/$q$, an enlarged magnetic unit cell $q$-times larger than the moiré unit cell (Fig. 2a) folds each electronic band into a reduced magnetic Brillouin zone $q$-times smaller than the moiré Brillouin zone, fractionalizing each band into $q$ Hofstadter subbands (Fig. 2b,c). Figs. 2d-g show the zero-bias conductance d$I$/d$V$($V_s$ = 0, $V_g$) at $B$ = 1.5 T, 1.75 T, 2.25 T, 3 T ($\Phi/\Phi_0 \approx$ 1/6, 1/5, 1/4, 1/3). As seen at zero $B$-field (Fig. 1g), deep suppressions of d$I$/d$V$($V_s$ = 0, $V_g$) appear at $v$ = ±4 at each $B$-field.

Additionally, each band splits into $q$ Hofstadter subbands, with new gaps between Hofstadter subbands at $v = \pm 4/q$ that reflect $\Phi/\Phi_0 = 1/q$. This parallels similar insulating states reported in transport measurements of graphene aligned to hBN at much higher $B$-fields ($B > 10$ T)[8–10].

The power of STS is its ability to resolve spectroscopic features both at and away from $E_F$, elucidating the Hofstadter spectrum's fractal structure and uncovering correlation effects that modify it. However, quasiparticle lifetime (QPL) effects, which affect all spectroscopic techniques, appear to be quite strong in second magic-angle TBG. Such effects present important challenges for realistic spectroscopic measurements unanticipated by naïve expectations of how experiments might access the Hofstadter spectrum. Fig. 2h illustrates the strong QPL effects in TBG, which limit the STM's ability to resolve LDOS features at arbitrarily large energies (red and blue spectrum regions). Instead, the STM's high spectroscopic resolution appears only effective within a narrow ±10 meV window surrounding $E_F$.

Strong QPL broadening effects can arise from a variety of mechanisms, notably from electron-electron interactions that are greatly enhanced near the second magic angle. Fortunately, the density-tunability of TBG enables us to resolve each Hofstadter subband using the STM's full energy resolution by doping subbands near $E_F$. Fig. 2i shows $dI/dV(V_s)$ at the same $B$-field, for a density of TBG chosen to bring broadened Hofstadter subbands close to $E_F$. In these spectra, Hofstadter subbands appear as sharp distinguishable LDOS peaks, illustrating the necessity of performing STS while controlling both the energy and density of the system.

Figs. 3a,c show $dI/dV(V_s, V_g)$ at $\Phi/\Phi_0 \approx 1/6$ and $1/3$ (see SI for other rational $\Phi/\Phi_0$ data). A few key features of our data enable us to identify a pair of topological invariants associated with each spectroscopic gap. First, groups of Hofstadter gaps emanate from fillings $v = 4N$ for integer $N$ at zero $B$-field, where the system's unbound charge density resets. Thus, Hofstadter gaps emanating from these special fillings differ by the bound-charge-associated topological invariant $s = \partial N_e/\partial N_{UC}$, where $N_e$ and $N_{UC}$ are the number of electrons in the gap in the zero $B$-field limit and the number of moiré unit cells, respectively[31]. By measuring the density $N_e/N_{UC}$ at which each Hofstadter gap occurs in STS as a function of $B$-field $N_\Phi/N_{UC}$ (where $N_\Phi$ is the number of magnetic flux quanta), STS measures its Chern number via Streda's formula[3,32]. Thus, Hofstadter gaps emanating from the same integer filling $s$ differ by the free-charge-associated topological invariant $t = \partial N_e/\partial N_\Phi$ (Chern number). Hofstadter gaps with Chern numbers $t = 4N$ for integer $N$ emanating from the same filling are evenly spaced in density. Therefore, we label each Hofstadter gap (right edges of Fig. 3a-d) by an ordered pair ($t$, $s$) of topological invariants, which appears at filling $v = t\Phi/\Phi_0 + s$ when seen at $E_F$ in STS data.

The zeroth Landau level (ZLL) / Chern $t = 0$ gaps, labeled (0, 4$N$) for integer $N$, appear at the same density at all $B$-fields, consistent with previous transport studies[12]. Furthermore, most Hofstadter gaps are unambiguously identified by their $B$-field dependence, particularly as observed away from simple $\Phi/\Phi_0$ ratios (see SI), but a few ambiguous cases remain with two possible ($t$, $s$) pairs. By tracking the bands associated with each zero-bias conductance peak in Figs. 2d-g and by using the invariants of their neighboring gaps, we identify each Hofstadter subband (dark labels between $V_g = 0$ V and -15 V; Figs. 3a,c) as a function of energy. Some Hofstadter bands move relatively rigidly to larger (smaller) energies when decreasing (increasing) $V_g$, reflecting the predicted single-particle Hofstadter spectrum. However, this picture breaks down in a few density regions. Near $v = 0$, correlation-driven gaps appear at partial fillings of the ZLL, which we analyze later. Additionally, certain gaps near $v = \pm 4$ anomalously vanish at moderate B-fields (B ≈ 3 T). For example, the (4, 4) gap closes, vanishing from the Hofstadter spectrum (Fig. 3c; $V_g \approx 22$ V) despite clear observations of neighboring (8, 4), (0, 4), and (-4, 4) gaps persisting.

To explain this, we construct a model based on the standard Bistritzer-MacDonald model for TBG with a minimal set of modifications, and use this model to calculate the LDOS[33] (Figs. 3b,d) for the twist angle and strain parameters of our devices (Figs. 3a,c). Our model includes interlayer heterostrain effects[34], which modify the Hofstadter spectrum in the absence of interactions. Additionally, we account for electron-electron Coulomb repulsion using a filling-dependent Hartree potential (see SI for details). LDOS calculations using our model (Figs. 3b,d; see SI for more comparisons) show excellent agreement with d$I$/d$V$($V_s$, $V_g$) at each $B$-field, capturing the density-dependent energetic evolution of Hofstadter bands, and notably reproducing the anomalously vanishing Hofstadter gaps. Calculations lacking interaction effects show weaker connections to STS data (see SI), suggesting our simple model is a faithful description of this system.

**Self-Similar Fractal Hofstadter Spectrum**

The defining feature of Hofstadter's spectrum is its fractal structure, which manifests as the spectrum's self-similar characteristics in energy and magnetic field[1,2,13]. In second magic-angle TBG, the Hofstadter spectrum can be approximated near $v = \pm 4$ and at low $B$-fields by single-particle calculations, and is predicted to exhibit self-similar energy distributions of Hofstadter subbands at $\Phi/\Phi_0 = 1/q$. Between these simple $\Phi/\Phi_0$ ratios, Hofstadter's spectrum exhibits additional complexities in its subband and gap structures that are inequivalent to those at $\Phi/\Phi_0 = 1/q$. We refer to this property as "discrete self-similarity", as the conditions for this scale invariance occur when $\Phi_0/\Phi$ is integer-valued (see SI for details).

We demonstrate signatures of discrete self-similarity by identifying scaling transformations of our STS data at simple fractional ratios of $\Phi/\Phi_0 = 1/q$. Fig. 4a shows $dI/dV(V_s, V_g)$ near $v = +4$ at $\Phi/\Phi_0 \approx 1/6, 1/5, 1/4$ ($B = 1.5$ T, 1.75 T, 2.25 T). The top row displays raw data obtained at each magnetic field, while the bottom row shows data transformed by a discrete scaling of the energy and density by a $B$-field-dependent linear scaling factor of $\frac{1}{4}(\Phi_0/\Phi)$, where $\Phi_0/\Phi = q$ is integer-valued (see SI for details). Similarly, Fig. 4c shows $dI/dV(V_s, V_g)$ near $v = -4$, scaled by the same factors.

Discrete scaling transformations yield STS spectra that resemble one another to a remarkable degree near $v = \pm 4$ (bottom panels; Figs. 4a,c). The densities of Hofstadter gaps[3] and the energies of Hofstadter subbands[35] are both expected to scale linearly with $B$-field when derived from quasi-parabolic band dispersions. However, the discrete nature of this self-similarity (i.e. by only integer-valued $\Phi_0/\Phi$) reflects a universal and distinctive feature of fractals[36,37], and is exhibited by Hofstadter's butterfly because of its recursive construction[1]. Thus, our observation of discrete self-similarity is a hallmark of the fractal nature of the Hofstadter spectrum.

Crucially, discrete self-similarity violations are predicted at non-integer $\Phi_0/\Phi$. Indeed, $dI/dV(V_s, V_g)$ data at non-integer $\Phi_0/\Phi$ ($\Phi/\Phi_0 \approx 5/18$; Figs. 4b,d) do not show discrete self-similarity. Detailed LDOS calculations of the Hofstadter spectrum (see SI) reflect these observations qualitatively, but deviate from STS data near regions affected by negative compressibility behavior. For example, negative compressibility appears near $v = -4$ as a backwards-dispersing Hofstadter band near $\frac{1}{4}(\Phi_0/\Phi)V_s = 5$ mV (red feature, lower panels of Fig. 4c; see SI). These discrepancies emphasize the role of electronic correlations in modifying the single-particle Hofstadter spectrum in ways difficult to capture fully with our current model, even at low $B$-fields.

**Interactions Alter Hofstadter's Spectrum**

Fig. 5a-c shows $dI/dV(V_s, B)$ at $v = +4, 0, -4$ (see supplementary video for full density-dependent data). These measurements reveal new correlation-driven effects on the Hofstadter spectrum beyond what is easily seen at fixed $B$-fields. For example, one might expect the bandwidths of Hofstadter subbands and the sizes of Hofstadter gaps to be density-independent, according to Hofstadter's calculations that ignore interaction effects[1]. Instead, STS measurements identify that the Hofstadter spectrum of TBG evolves dynamically with density

(see supplementary video). Hofstadter gaps appear in the spectrum at some densities and disappear at others (e.g. $v = 0$ gap in Fig. 5b vs. Figs. 5a,c). Moreover, some Hofstadter subbands split when doped to $E_F$, as discussed next. All of these effects result from interactions.

Finally, we observe an unusual dichotomy between the effects of strong correlations at low B-fields that disappear at higher B-fields and those that arise only at higher B-fields (Figs. 3c,d). Fig. 5d plots $dI/dV(V_s)$ spectra of the $(t, s) = (4, 0), (-4, 0)$ gaps within the ZLL (black triangle markers) at three B-fields. The ZLL increases dramatically in bandwidth and additionally splits into two well-separated peaks with increasing B-field (see SI for bandwidth analysis). The ZLL can be alternatively understood as the two Hofstadter subbands closest to charge neutrality. As B increases, flat bands in TBG split into fewer and fewer Hofstadter subbands that become broader and broader, a direct consequence of band folding effects predicted by Hofstadter's calculations[1], thus leading to their increased bandwidth at higher B-fields.

$dI/dV(V_s,B)$ and $dI/dV(V_s,V_g)$ (Figs. 5e,f) both illustrate these Hofstadter-band broadening effects. At low B-fields ($B < 2$ T), a correlation-driven exchange gap opens at $E_F$ at partial fillings of the ZLL (Fig. 5b)[38,30]. Increasing the B-field reduces the exchange gap, which is ultimately replaced by a peak in LDOS at $B = 3$ T. Moreover, at $B = 1.5$ T ($\Phi/\Phi_0 \approx 1/6$; bottom panel of Fig. 5c), an 8-fold-degenerate ZLL appears as a narrow pair of LDOS peaks (e.g. $V_s = -8$ mV, $V_g = 8$ V), which splits at $E_F$ at densities of $v_{LL}B/\Phi_0$ for Landau level fillings $v_{LL} = -2, 0, +2$ due to quantum Hall ferromagnetism[39,40,29]. With increasing B-field, these correlation-driven insulating gaps (yellow triangle markers) and the exchange gap (white triangle markers) both weaken, disappearing by $B = 3$ T.

This behavior is unlike quantum Hall ferromagnetism in other systems, which is typically absent at lower B-fields and enhanced at higher B-fields due to stronger correlation effects from increased Landau-level degeneracy and interaction strength. However, in Hofstadter systems, concomitant with increasing interaction strength is a broadening of Hofstadter subbands as they merge at high B-fields. Hofstadter-band broadening effects dominate in second magic-angle TBG, which regains its full isospin symmetry (top panel; Fig. 5f) as B increases (see SI for discussion). Under very different experimental conditions, a likely related effect was noticed in compressibility experiments on graphene aligned to hBN at very high B-fields ($\Phi/\Phi_0 > 1$)[14]. However, its origin was elusive. Here, spectroscopy identifies this feature as arising from Hofstadter-band broadening effects.

Our measurements uncover spectroscopic signatures of the Hofstadter spectrum in second magic-angle TBG, a model platform for high-resolution measurements of its fractal band structure. We expect that future studies may use the Hofstadter spectrum as a sensitive probe of topology and correlation effects in moiré materials. As we have demonstrated, both effects imprint themselves directly upon the system's Hofstadter spectrum in ways more obvious than those observable at zero *B*-field[35,41]. Such information is crucial for isolating minimal descriptions of moiré materials that capture the salient electronic phenomena in these complex settings.

**References:**

1. Hofstadter, D. R. Energy levels and wave functions of Bloch electrons in rational and irrational magnetic fields. *Phys. Rev. B* **14**, 2239–2249 (1976).

2. Wannier, G. H. A Result Not Dependent on Rationality for Bloch Electrons in a Magnetic Field. *Phys. Status Solidi B* **88**, 757–765 (1978).

3. Streda, P. Quantised Hall effect in a two-dimensional periodic potential. *J. Phys. C Solid State Phys.* **15**, L1299 (1982).

4. Gerhardts, R. R., Weiss, D. & Wulf, U. Magnetoresistance oscillations in a grid potential: Indication of a Hofstadter-type energy spectrum. *Phys. Rev. B* **43**, 5192–5195 (1991).

5. Albrecht, C. *et al.* Evidence of Hofstadter's Fractal Energy Spectrum in the Quantized Hall Conductance. *Phys. Rev. Lett.* **86**, 147–150 (2001).

6. Geisler, M. C. *et al.* Detection of a Landau Band-Coupling-Induced Rearrangement of the Hofstadter Butterfly. *Phys. Rev. Lett.* **92**, 256801 (2004).

7. Yankowitz, M. *et al.* Emergence of superlattice Dirac points in graphene on hexagonal boron nitride. *Nat. Phys.* **8**, 382–386 (2012).

8. Dean, C. R. *et al.* Hofstadter's butterfly and the fractal quantum Hall effect in moiré superlattices. *Nature* **497**, 598–602 (2013).

9. Hunt, B. *et al.* Massive Dirac Fermions and Hofstadter Butterfly in a van der Waals Heterostructure. *Science* **340**, 1427–1430 (2013).

10. Ponomarenko, L. A. *et al.* Cloning of Dirac fermions in graphene superlattices. *Nature* **497**, 594–597 (2013).

11. Bistritzer, R. & MacDonald, A. H. Moiré bands in twisted double-layer graphene. *Proc. Natl. Acad. Sci.* **108**, 12233–12237 (2011).

12. Lu, X. *et al.* Multiple flat bands and topological Hofstadter butterfly in twisted bilayer graphene close to the second magic angle. *Proc. Natl. Acad. Sci.* **118**, e2100006118 (2021).

13. Bistritzer, R. & MacDonald, A. H. Moir\'e butterflies in twisted bilayer graphene. *Phys. Rev. B* **84**, 035440 (2011).

14. Yu, G. L. *et al.* Hierarchy of Hofstadter states and replica quantum Hall ferromagnetism in graphene superlattices. *Nat. Phys.* **10**, 525–529 (2014).

15. Krishna Kumar, R. *et al.* High-temperature quantum oscillations caused by recurring Bloch states in graphene superlattices. *Science* **357**, 181–184 (2017).

16. Barrier, J. *et al.* Long-range ballistic transport of Brown-Zak fermions in graphene superlattices. *Nat. Commun.* **11**, 5756 (2020).

17. Yankowitz, M., Ma, Q., Jarillo-Herrero, P. & LeRoy, B. J. van der Waals heterostructures combining graphene and hexagonal boron nitride. *Nat. Rev. Phys.* **1**, 112–125 (2019).

18. Spanton, E. M. *et al.* Observation of fractional Chern insulators in a van der Waals heterostructure. *Science* **360**, 62–66 (2018).

19. Kometter, C. R. *et al.* Hofstadter states and reentrant charge order in a semiconductor moir\'e lattice. Preprint at https://doi.org/10.48550/arXiv.2212.05068 (2022).

20. Yoo, H. *et al.* Atomic and electronic reconstruction at the van der Waals interface in twisted bilayer graphene. *Nat. Mater.* **18**, 448–453 (2019).

21. Kazmierczak, N. P. *et al.* Strain fields in twisted bilayer graphene. *Nat. Mater.* **20**, 956–963 (2021).

22. Wong, D. *et al.* Cascade of electronic transitions in magic-angle twisted bilayer graphene. *Nature* **582**, 198–202 (2020).

23. Cao, Y. *et al.* Correlated insulator behaviour at half-filling in magic-angle graphene superlattices. *Nature* **556**, 80–84 (2018).

24. Cao, Y. *et al.* Unconventional superconductivity in magic-angle graphene superlattices. *Nature* **556**, 43–50 (2018).

25. Yankowitz, M. *et al.* Tuning superconductivity in twisted bilayer graphene. *Science* **363**, 1059–1064 (2019).

26. Lu, X. *et al.* Superconductors, orbital magnets and correlated states in magic-angle bilayer graphene. *Nature* **574**, 653–657 (2019).

27. Chan, H. B., Glicofridis, P. I., Ashoori, R. C. & Melloch, M. R. Universal Linear Density of States for Tunneling into the Two-Dimensional Electron Gas in a Magnetic Field. *Phys. Rev. Lett.* **79**, 2867–2870 (1997).

28. Eisenstein, J. P., Pfeiffer, L. N. & West, K. W. Coulomb barrier to tunneling between parallel two-dimensional electron systems. *Phys. Rev. Lett.* **69**, 3804–3807 (1992).

29. Liu, X. *et al.* Visualizing broken symmetry and topological defects in a quantum Hall ferromagnet. *Science* **375**, 321–326 (2022).

30. Hu, Y. *et al.* High-Resolution Tunneling Spectroscopy of Fractional Quantum Hall States. Preprint at https://doi.org/10.48550/arXiv.2308.05789 (2023).

31. MacDonald, A. H. Landau-level subband structure of electrons on a square lattice. *Phys. Rev. B* **28**, 6713–6717 (1983).

32. Nuckolls, K. P. *et al.* Strongly correlated Chern insulators in magic-angle twisted bilayer graphene. *Nature* **588**, 610–615 (2020).

33. Michael G. Scheer, Jonah Herzog-Arbeitman, Kevin P. Nuckolls, Ali Yazdani, & Biao Lian. Hofstadter Band Theory for Continuum Models. *in preparation* (2024).

34. Bi, Z., Yuan, N. F. Q. & Fu, L. Designing flat bands by strain. *Phys. Rev. B* **100**, 035448 (2019).

35. Lian, B., Xie, F. & Bernevig, B. A. Landau level of fragile topology. *Phys. Rev. B* **102**, 041402 (2020).

36. Mandelbrot, B. How Long Is the Coast of Britain? Statistical Self-Similarity and Fractional Dimension. *Science* **156**, 636–638 (1967).

37. Falconer, K. *Fractal Geometry: Mathematical Foundations and Applications*. (John Wiley & Sons, 2014).

38. Dial, O. E., Ashoori, R. C., Pfeiffer, L. N. & West, K. W. High-resolution spectroscopy of two-dimensional electron systems. *Nature* **448**, 176–179 (2007).

39. Nomura, K. & MacDonald, A. H. Quantum Hall Ferromagnetism in Graphene. *Phys. Rev. Lett.* **96**, 256602 (2006).

40. Young, A. F. *et al.* Spin and valley quantum Hall ferromagnetism in graphene. *Nat. Phys.* **8**, 550–556 (2012).

41. Herzog-Arbeitman, J., Song, Z.-D., Regnault, N. & Bernevig, B. A. Hofstadter Topology: Noncrystalline Topological Materials at High Flux. *Phys. Rev. Lett.* **125**, 236804 (2020).

42. Wong, D. *et al.* A modular ultra-high vacuum millikelvin scanning tunneling microscope. *Rev. Sci. Instrum.* **91**, 023703 (2020).

**Acknowledgements**

We thank B. Andrei Bernevig for useful discussions. This experimental work reported here was primarily supported by the DOE-BES grant DE-FG02-07ER46419 to A.Y. Support for the theory-experimental collaboration reported here was provided by NSF-MRSEC through the Princeton Center for Complex Materials grant NSFDMR-2011750 to A.Y. and B.L. Additional support for the experimental work was provided by the Gordon and Betty Moore Foundation's EPiQS initiative grant GBMF9469 and the U.S. Army Research Office MURI project under grant number W911NF-21-2-0147, ONR grant N000142412471, and NSF grant DMR-2312311 to A.Y. M. S. and B. L. are supported by the National Science Foundation under award DMR-2141966. Additional support is provided by the Gordon and Betty Moore Foundation through Grant GBMF8685 towards the Princeton theory program. J.H.A. was supported by a Hertz Fellowship. K.P.N. acknowledges support from the MIT Pappalardo Fellowship in Physics during the preparation of this manuscript. K.W. and T.T. acknowledge support from the Elemental Strategy Initiative conducted by the MEXT, Japan, grant JPMXP0112101001, JSPS KAKENHI grant 19H05790 and JP20H00354.

**Author Contributions**

K.P.N., D.W., M.O., and A.Y. designed the experiment. K.P.N. fabricated the devices used for the study. K.P.N., D.W., and M.O. carried out STM / STS measurements. M.S. and J.H.A. constructed the electronic structure model and performed local density of states simulations,

## Methods

### STM / STS measurements

STM / STS measurements were performed using a homebuilt, ultrahigh vacuum (UHV) STM[42] either at $T$ = 4 K or $T$ = 200 mK. Tungsten STM tips were prepared on a Cu(111) single crystal and calibrated against the crystal's Shockley surface state. Twisted bilayer graphene was electrostatically gated through a voltage $V_g$ applied to a degenerately p-doped Si back-gate, and was biased through a voltage $V_s$ applied to the sample relative to the tip. Scanning tunnelling spectroscopy (STS) measurements were obtained through a standard lock-in detection method of the tunnelling current, induced by a sinusoidal modulation voltage added to the sample bias. See SI for tunneling parameters.

### Sample preparation.

Devices were fabricated using a 'tear-and-stack' method using a method similar to the one described in detail in Ref. [22]. Briefly, graphene and hBN were picked up using a polyvinyl alcohol (PVA) film handle, flipped onto an intermediate Sylgard 184 handle, and dropped off onto a prepatterned $SiO_2$ chip with Ti / Au electrodes. Residual polymer was dissolved using a combination of N-methyl-2-pyrrolidone, water, acetone, and isopropyl alcohol. The device surface was further cleaned using atomic-force-microscope-tip cleaning and ultrahigh vacuum annealing first at 170 °C for 12 hr, followed by an anneal at 400 °C for 3 hr.

### Data Availability

The data that supports the findings of this study are available from the corresponding author upon reasonable request.

### Code Availability

The code that supports the findings of this study is available as an open-source Compute Capsule supported by a partnership between Code Ocean and Springer Nature.

### Competing Interests

The authors declare no competing interests.

# Figure 1

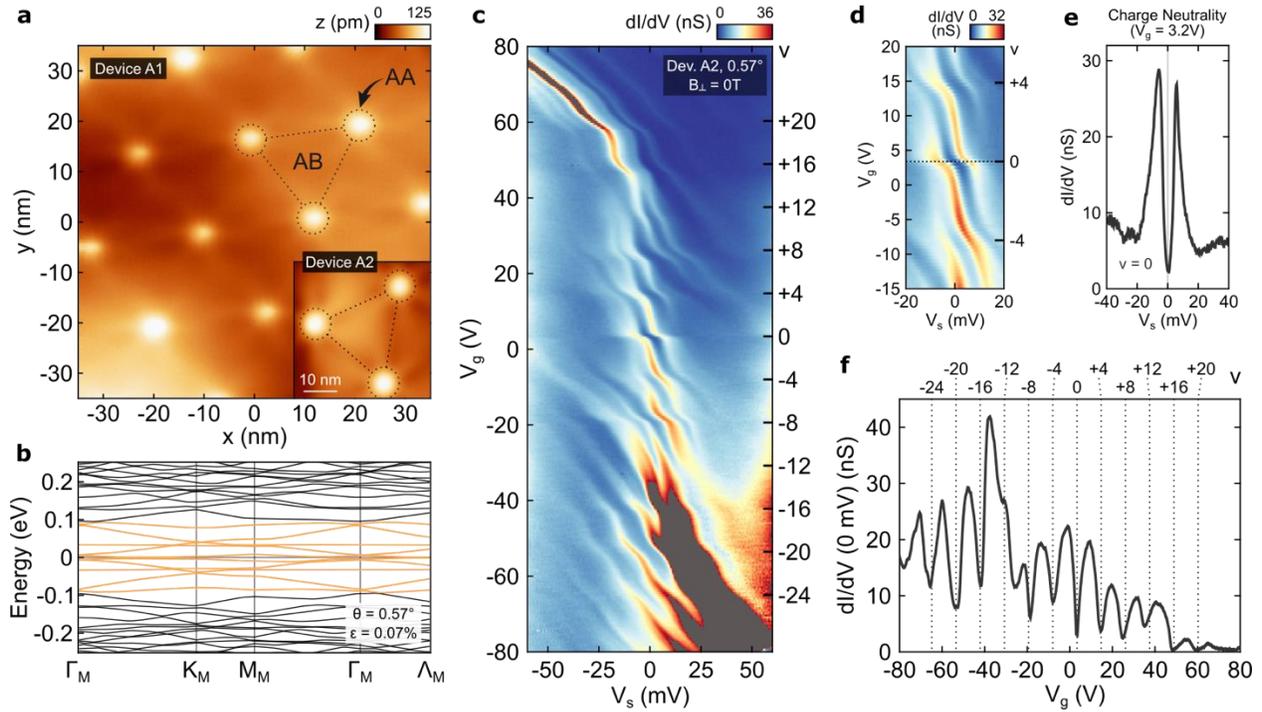

**Figure 1 | Abundance of Flat Bands in Twisted Bilayer Graphene near the 2nd Magic Angle. a**, STM topographic images of twisted bilayer graphene (TBG) near the second magic angle (Device A1: $\theta = 0.63°$, $\varepsilon = 0.04\%$; Device A2: $\theta = 0.57°$, $\varepsilon = 0.07\%$). Inset z-scale: 0 pm to 135 pm. **b,** Electronic band structure calculation for TBG using Device A2 twist and strain parameters. Near the Fermi energy ($E_F$), single-particle calculations predict a dense set of flat electronic bands. The 10 flat bands nearest to $E_F$ are labeled in orange. **c,** d$I$/d$V(V_s, V_g)$ measured at the center of an AA site at $T = 4.1$ K and $B = 0$ T in an ultra-low strain region of TBG (Device A2). Sharp peaks in the local density of states (LDOS) each pin to the Fermi level, separated by deep suppressions / gaps in LDOS at integer fillings $v = 4N$ for integer $N$. **d,** Zoom-in plot of the flat bands near charge neutrality in **c**. Two sharp peaks appear in LDOS, comparable in bandwidth to those observed in TBG near the first magic angle. **e,** d$I$/d$V(V_s)$ obtained at charge neutrality. **f,** Zero-bias conductance d$I$/d$V(V_s = 0, V_g)$ obtained from a line cut of **c**, showing deep conductance suppressions at twelve fillings $v = 4N$ for integer $N$. See SI for tunneling parameters.

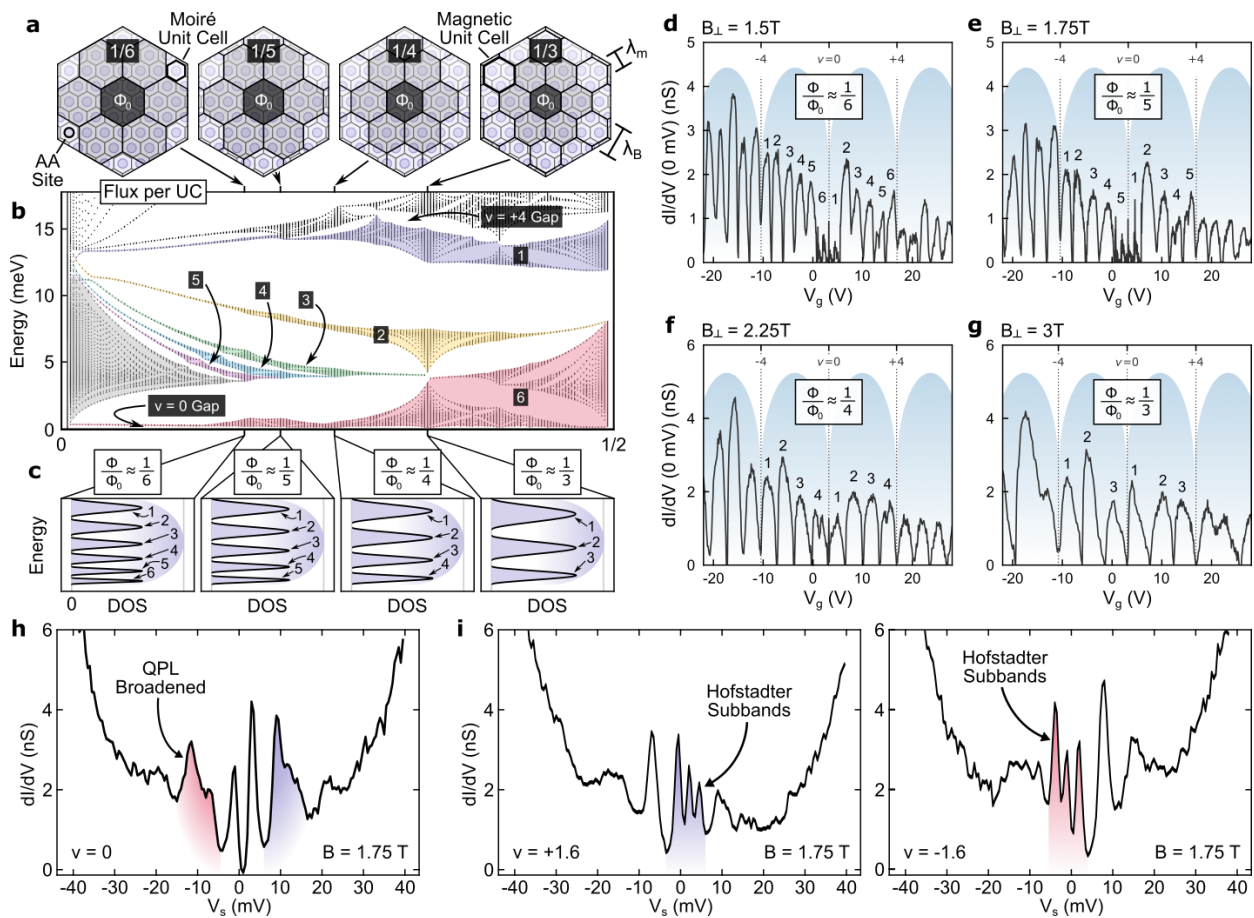

**Figure 2 | Hofstadter's Butterfly in Twisted Bilayer Graphene near Rational Flux Ratios. a,** Schematic diagrams of the magnetic unit cell of TBG as a function of magnetic field. At rational values of magnetic flux per moiré unit cell $\Phi/\Phi_0 = 1/q$, the magnetic unit cell (with lattice constant $\lambda_B$) is enlarged by a factor of $q$ compared to the moiré unit cell (with lattice constant $\lambda_m$). **b,** Calculated Hofstadter spectrum of TBG[33] near the second magic angle (Device A1: $\theta$ = 0.63°, $\varepsilon$ = 0.04%). At $\Phi/\Phi_0 = 1/q$, each flat band of TBG reorganizes into $q$ Hofstadter subbands. **c,** Schematic diagrams of the density of states expected at rational $\Phi/\Phi_0$, according to the Hofstadter subbands in **b**. Numerical labels count the number of dominant Hofstadter subbands seen in spectroscopy. **d-g,** Zero-bias tunneling conductance $dI/dV(V_s = 0, V_g)$ obtained at the center of an AA site at four rational values of magnetic flux per unit cell in Device A1. Peaks in the zero-bias conductance parallel the Hofstadter subbands shown in **b**. Hofstadter bands close to charge neutrality further split in **d** and **e** due to quantum Hall ferromagnetism shown in Fig. 5. **h**, Tunneling spectroscopy $dI/dV(V_s)$ at charge neutrality ($v = 0$, $V_g = 3$ V) at $B = 1.75$ T ($\Phi/\Phi_0 \approx$ 1/5). Hofstadter subbands near $E_F$ ($V_s = 0$) appear sharp, but those away from $E_F$ (red and blue

spectrum regions) appear broadened by strong quasiparticle lifetime (QPL) effects. **i**, Same as **h**, at fillings $v = \pm1.6$ ($V_g = 5.6$ V, -2.6 V), where higher energy Hofstadter subbands (red and blue spectrum regions) can be resolved when appearing near $E_F$. See SI for tunneling parameters.

# Figure 3

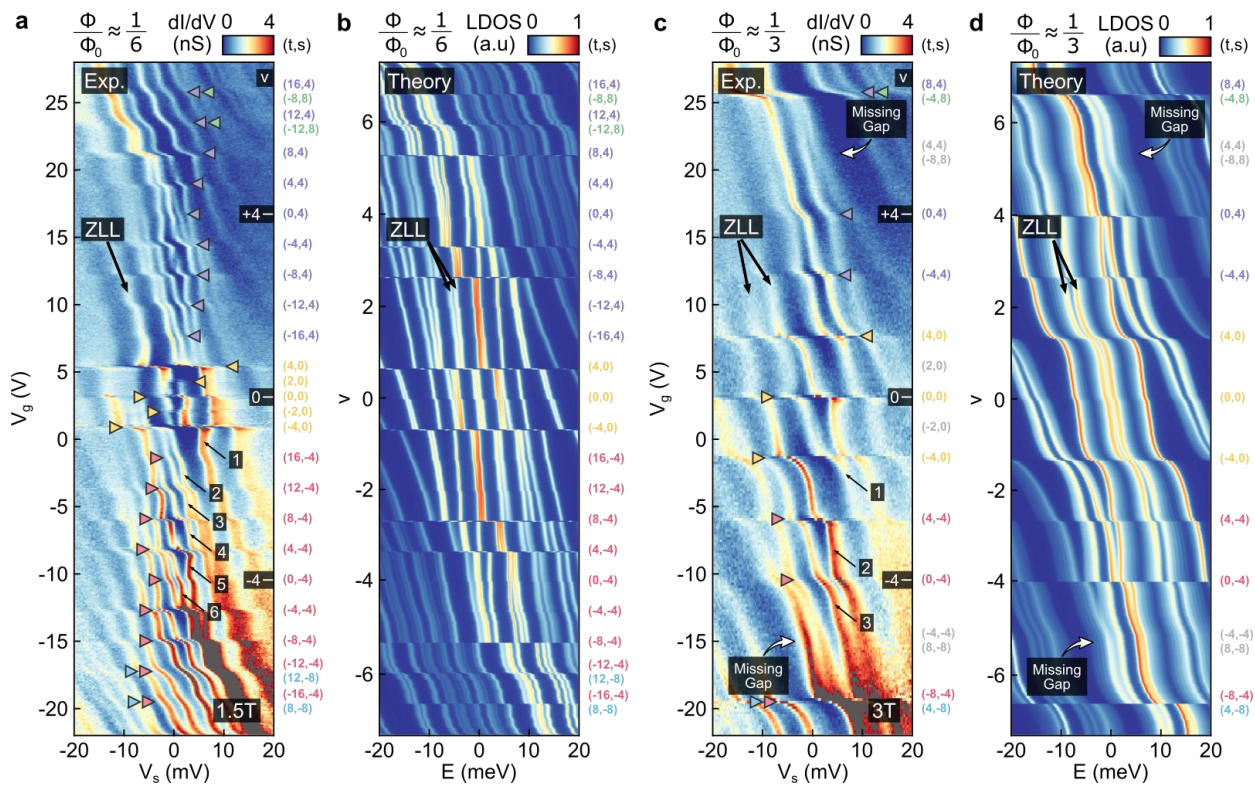

**Figure 3 | Direct Spectroscopic Resolution of the Fractal Hofstadter Spectrum. a,** d$I$/d$V$($V_s$, $V_g$) obtained at the center of an AA site in Device A1 at $B$ = 1.5 T ($\Phi/\Phi_0 \approx 1/6$), where each flat band fractionalizes into 6 Hofstadter subbands (dark labels between $v$ = 0 and $v$ = -4). Hofstadter gaps emanating from fillings $v$ = +8, +4, 0, -4, -8 are labeled by green, purple, yellow, red, and blue triangular markers, respectively. **b,** Calculated local density of states of TBG[33] at $\Phi/\Phi_0$ = 1/6. **c,** Same as **a**, for $B$ = 3 T ($\Phi/\Phi_0 \approx 1/3$). **d,** Same as **b**, for $\Phi/\Phi_0$ = 1/3. See SI for tunneling parameters.

# Figure 4

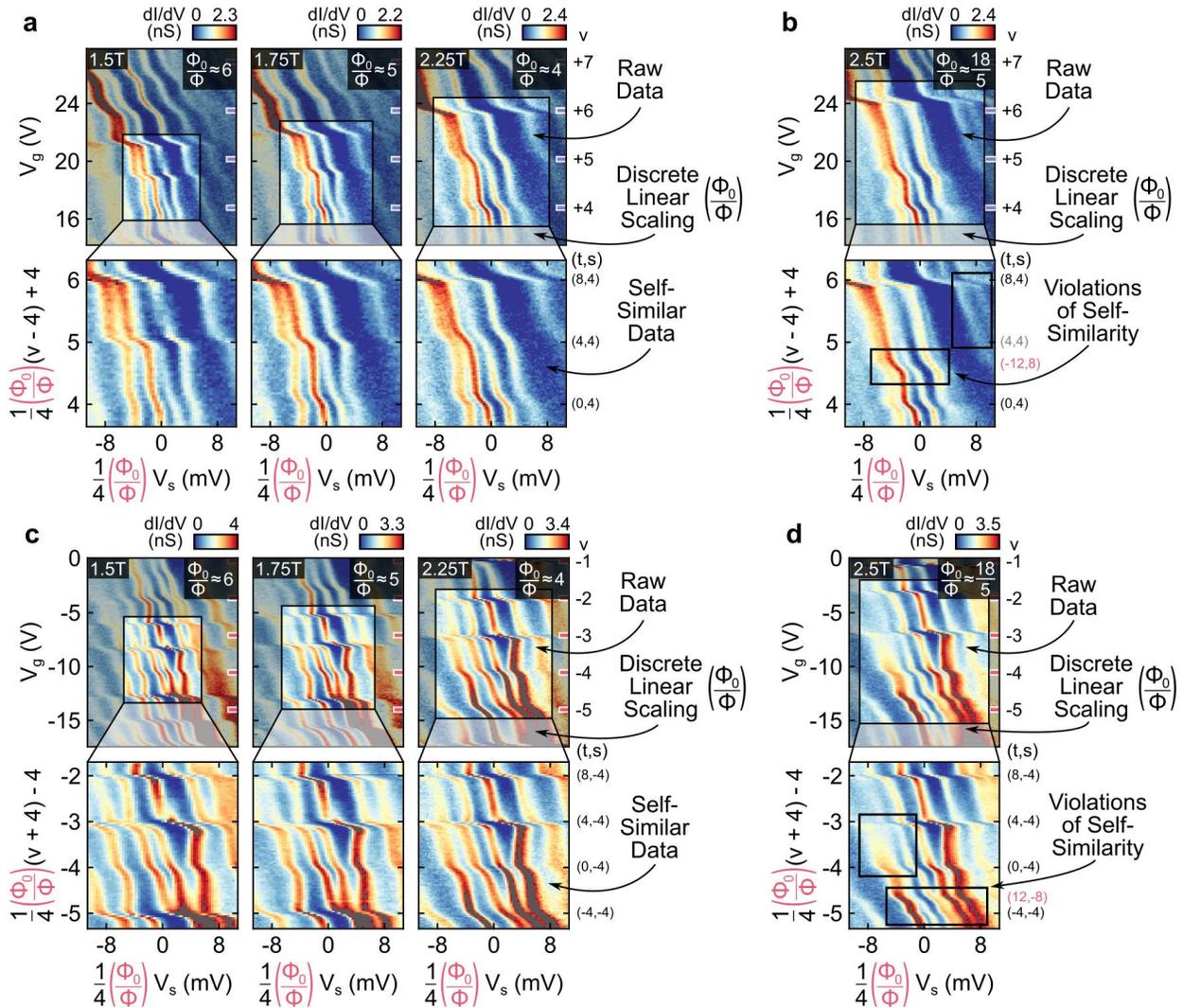

**Figure 4 | Self-Similarity of the Fractal Hofstadter Spectrum. a,** d$I$/d$V$($V_s$, $V_g$) obtained at the center of an AA site in Device A1 near $v$ = +4 at $B$ = 1.5 T (Φ/Φ$_0$ ≈ 1/6; left), 1.75 T (Φ/Φ$_0$ ≈ 1/5; middle), and 2.25 T (Φ/Φ$_0$ ≈ 1/4; right). The top row of panels shows raw d$I$/d$V$($V_s$, $V_g$) (horizontal axes range from -13.5 mV to 13.5 mV). The bottom row shows data scaled in energy and density by a discrete linear scaling factor of ¼(Φ$_0$/Φ) where Φ$_0$/Φ is an integer. The full density expression ¼(Φ$_0$/Φ)($v$ - $v_{ref}$) + $v_{ref}$, where $v_{ref}$ = +4, both scales the data and further vertically aligns the scaled data. **b,** Same as **a**, at $B$ = 2.5 T (Φ/Φ$_0$ ≈ 5/18). Violations of discrete self-similarity are highlighted in black outlined boxes. **c,** Same as **a**, near $v$ = -4. The full density expression is the same as in **a**, with $v_{ref}$ = -4. **d,** Same as **c**, at $B$ = 2.5 T (Φ/Φ$_0$ ≈ 5/18). Violations of discrete self-similarity are highlighted in black outlined boxes. See SI for tunneling parameters.

# Figure 5

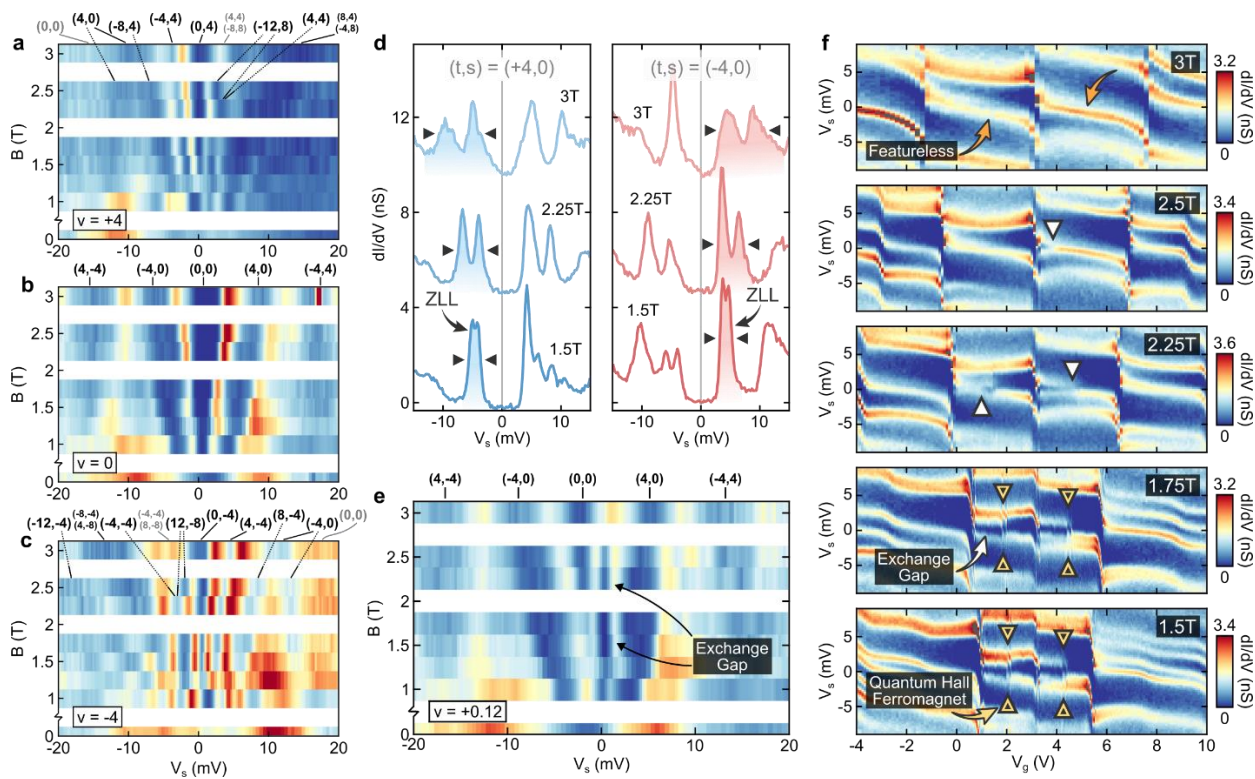

**Figure 5 | Correlation-Driven Evolution of the Hofstadter Spectrum. a-c,** $dI/dV(V_s,B)$ at fixed gate voltages $V_g$ = 16.8 V (**a**; $\nu$ = +4), 3 V (**b**; $\nu$ = 0), and -10.8 V (**c**; $\nu$ = -4). Black number labels identify the topological invariants of gaps observed in the Hofstadter spectrum; whereas, gray number labels identify Hofstadter gaps that were previously observed at a different density and magnetic field and have vanished from the spectrum. **d,** $dI/dV(V_s)$ spectra obtained for the Chern +4 (left; $V_g$ = 7.6 V, 6.5 V, 5.4 V) and Chern -4 (right; $V_g$ = -1.4 V, -0.2 V, 0.9 V) zeroth Landau level (ZLL) gaps near charge neutrality at three $B$-field values. As the $B$-field increases, the ZLL (black arrow markers) splits and broadens in energy. Curves offset for clarity. **e,** $dI/dV(V_s,B)$ at a fixed $V_g$ = 3.4 V ($\nu$ = +0.12). The interaction driven exchange gap, which only appears at the Fermi level ($V_s$ = 0), weakens with increasing $B$, finally vanishing at $B$ = 3 T. **f,** $dI/dV(V_s, V_g)$ of the ZLL obtained at five $B$-fields. As the $B$-field increases, signatures of electronic interactions, including quantum-Hall-ferromagnetic gaps (yellow arrow markers) and the exchange gap (white arrow markers) vanish. See SI for color bar details (**a-c**) and for tunneling parameters.